\documentclass[a4paper]{llncs}

\usepackage{hyperref}
\usepackage{algorithmic}
\usepackage{float}
\usepackage[font=small,skip=0pt]{caption}
\usepackage[font=footnotesize]{subcaption}
\usepackage{listings}
\usepackage{lstcoq}
\usepackage{todonotes}
\usepackage{amsmath,amssymb,amsfonts,mathtools}

\usepackage{graphicx,epsfig}
\usepackage{algorithm2e}
\usepackage{ifthen,array}
\usepackage{verbatim,alltt,color,multirow}

\usepackage{enumitem}
\usepackage{textcomp}
\usepackage{stmaryrd}

\let\llncssubparagraph\subparagraph
\let\subparagraph\paragraph
\usepackage{titlesec}
\let\subparagraph\llncssubparagraph

\newcommand{\var}{\mathrm{var}}

\newcommand{\fml}[1]{{\mathcal{#1}}}

\newcommand{\nfmt}{GRIT\xspace}

\DeclareMathOperator*{\entails}{\vDash}
\DeclareMathOperator*{\props}{\vdash}
\DeclareMathOperator*{\uprop}{\vdash_{\mathllap{\scalebox{0.575}{$u$}}}}

\DeclareMathSymbol{\Delta}{\mathalpha}{operators}{1}
\DeclareMathSymbol{\Theta}{\mathalpha}{operators}{2}
\DeclareMathSymbol{\Pi}{\mathalpha}{operators}{5}
\DeclareMathSymbol{\Sigma}{\mathalpha}{operators}{6}

\definecolor{gray}{rgb}{.4,.4,.4}
\definecolor{midgrey}{rgb}{0.5,0.5,0.5}
\definecolor{darkgrey}{rgb}{0.125,0.125,0.125}
\definecolor{darkred}{rgb}{0.7,0.1,0.1}
\definecolor{darkblue}{rgb}{0.1,0.1,0.5}
\definecolor{darkyellow}{rgb}{0.5,0.5,0.1}
\definecolor{darkgreen}{rgb}{0.1,0.5,0.1}
\definecolor{defseagreen}{cmyk}{0.69,0,0.50,0}

\newcommand{\jnotef}[1]{}
\newcounter{Comment}[Comment]
\setitemize{noitemsep,topsep=3pt,parsep=2pt,partopsep=2pt}
\setenumerate{noitemsep,topsep=3pt,parsep=2pt,partopsep=2pt}

\titlespacing{\section}{0pt}{*2.5}{*1.0}
\titlespacing{\subsection}{0pt}{*1.5}{*0.75}
\titlespacing{\subsubsection}{0pt}{*1.0}{*0.5}

\newcolumntype{L}[1]{>{\raggedright\let\newline\\\arraybackslash\hspace{0pt}}m{#1}}
\newcolumntype{C}[1]{>{\centering\let\newline\\\arraybackslash\hspace{0pt}}m{#1}}
\newcolumntype{R}[1]{>{\raggedleft\let\newline\\\arraybackslash\hspace{0pt}}m{#1}}

\SetAlFnt{\small}
\SetAlCapFnt{\small}
\SetAlCapNameFnt{\small}
\algsetup{linenosize=\tiny}

\hypersetup{
    bookmarks=false,
    pdftitle={},
    pdfauthor={Joao Marques-Silva},
    pdfsubject={TeX and LaTeX},
    pdfkeywords={TeX, LaTeX, graphics, images},
    colorlinks=true,
    linkcolor=blue,
    citecolor=blue,
    filecolor=black,
    urlcolor=blue,
    linktoc=page
}

\AtBeginDocument{

}

\definecolor{dkviolet}{rgb}{.5,0,.5}
\definecolor{dkblue}{rgb}{0,0,.5}
\definecolor{dkgreen}{rgb}{0,.5,0}
\definecolor{dkred}{rgb}{.5,0,0}
\definecolor{ltblue}{rgb}{0,.8,1}

\begin{document}

\title{
  Efficient Certified Resolution Proof Checking
}

\author{
  Lu\'\i s Cruz-Filipe\inst{1} \and
  Joao Marques-Silva\inst{2} \and
  Peter Schneider-Kamp\inst{1}}
\authorrunning{L.~Cruz-Filipe, J.~Marques-Silva, P.~Schneider-Kamp}
\institute{Department of Mathematics and Computer Science,\\
  University of Southern Denmark, Odense, Denmark\\
  \email{$\{$lcf,petersk$\}$@imada.sdu.dk}
  \and
  LaSIGE, Faculty of Science, University of Lisbon, Lisbon, Portugal\\
  \email{jpms@ciencias.ulisboa.pt}
}

\maketitle
\setcounter{footnote}{0}

\begin{abstract}
  We present a novel propositional proof tracing format that 
  eliminates complex processing, thus enabling efficient (formal) proof checking.
  The benefits of this format are demonstrated by implementing a proof
  checker in C, which outperforms a state-of-the-art checker by two
  orders of magnitude.
  We then formalize the theory underlying propositional proof checking
  in Coq, and extract a correct-by-construction proof checker for our
  format from the formalization.
  An empirical evaluation using 280 unsatisfiable instances from the
  2015 and 2016 SAT competitions shows that this certified checker
  usually performs comparably to a state-of-the-art non-certified proof checker.
  Using this format, we formally verify the recent $200$ TB proof of
  the Boolean Pythagorean Triples conjecture.
\end{abstract}

\section{Introduction} \label{sec:intro}

The practical success of Boolean Satisfiability (SAT) solvers cannot
be overstated. Generally accepted as a mostly academic curiosity until
the early 1990s, SAT solvers are now used ubiquitously, in a variety of
industrial settings, and with an ever increasing range of practical
applications~\cite{sat-handbook09}. Several of these applications are
safety-critical, and so in these cases it is essential that produced
results have some guarantee of correctness~\cite{shankar-atva08}.

One approach investigated over the years has been to develop
formally derived SAT
solvers~\cite{smith-tr08,maric-tcs10,maric-lmcs11,weidenbach-ijcar16}.
These works all follow the same underlying idea: formally specify SAT
solving techniques within a constructive theorem prover
and apply program extraction (an implementation of the Curry--Howard
correspondence) to obtain a certified SAT solver.
Unfortunately, certified SAT solvers produced by this method cannot
match the performance of carefully hand-optimized solvers, as these
optimizations typically rely on low-level code whose correctness is
extremely difficult to prove formally, and the performance gap is
still quite significant.

An alternative approach that has become quite popular is to \emph{check} the
results produced by SAT solvers, thus adding some level of assurance regarding
the computed results.
This line of work can be traced at least to the seminal work of Blum and
Kannan~\cite{blum-stoc89}, with recent work also focusing on certifying
algorithms and their verification~\cite{mehlhorn-csr11,mehlhorn-jar14}.
Most SAT checkers expect the SAT solver to produce a witness of its
result, and then validate the witness against the input formula. For
satisfiable instances, this is a trivial process that amounts to checking the
computed satisfying assignment against the input formula.
For unsatisfiable instances, since SAT is known to be in NP and believed not to
be in coNP, it is unlikely that there exist succinct witnesses, in the worst
case. As a result, the solution in practice
has been to output a \emph{trace} of the execution of the SAT
solver, which essentially captures a resolution proof of the
formula's unsatisfiability.
Although this approach finds widespread
use~\cite{goldberg-date03,malik-date03,biere-csr06,biere-sat06,biere-sat07,biere-jsat08,vangelder-isaim08,heule-cade13,heule-itp13,heule-fmcad13,heule-sat14,heule-stvr14,heule-sat14,heule-appa14,heule-fmcad14,heule-cade15,wetzler-phd15},
and has been used to check large-scale resolution
proofs~\cite{konev-corr14b,konev-sat14,selman-cp14,konev-aij15,heule-sat16a},
its main drawback is that there still is effectively no guarantee 
that the computed result is correct,
since the proof checker has again not been proven correct.

Combining these two approaches, several
authors~\cite{weber-jal09,dfms-coq09,dfms-ictac10,faure-cpp11,heule-itp13,heule-stvr14}
have experimented with the idea of developing \emph{certified} proof checkers,
i.e.~programs that check traces of unsatisfiability proofs and that have
themselves been formally proven correct.
However, all these approaches are limited in their scalability,
essentially for one of two reasons: (1) information about deletion of
learned clauses is not available nor
used~\cite{weber-jal09,dfms-coq09,dfms-ictac10,faure-cpp11}; and 
(2) the formats used to provide proof traces by SAT solvers still
require the checker to perform complex checking
steps~\cite{heule-cade13,heule-itp13,heule-stvr14,wetzler-phd15},
which are very difficult to optimize.

In this paper we examine the fundamental reasons for why these
attempts do not scale in practice, and propose
a resolution proof trace format that extends the one developed in recent
work~\cite{heule-cade13,heule-fmcad13,heule-sat14,heule-stvr14} by incorporating
enough information to allow the reconstruction of the original resolution proof
with minimum computational effort.
This novel proof trace format impacts resolution proof checking in a
number of fundamental aspects.
First, we show how we can implement an (uncertified, optimized) proof checker in
C whose run times are negligible when compared to those of state-of-the-art
checkers, in particular
drat-trim~\cite{heule-sat14,drat-trim}\footnote{The sole purpose of
  comparing two checkers with different aims and based on different
  formats is to motivate the development of the efficient certified
  checker.}.
Second, we capitalize on the simplicity of the new proof format to formalize the
proof verification algorithm inside the theorem prover Coq.
Third, we extract a certified checker from this formalization and show that it
performs comparably with drat-trim on a number of significant
test cases.
As a consequence, this certified checker is able to verify, in reasonable time, 
the currently largest available resolution proof, namely the $200$ TB proof of
the unsatisfiability of a SAT encoding of the Boolean Pythagorean Triples conjecture~\cite{heule-sat16a}.

The paper is organized as follows.
\autoref{sec:prelim} briefly summarizes basic SAT and proof checking
definitions, and presents a brief overview of the Coq theorem prover and its
extraction mechanism.
\autoref{sec:fmts} provides an overview of the best known resolution
proof formats proposed in the recent past.
\autoref{sec:nfmt} introduces the novel resolution proof trace format and
outlines the pseudo-code of a verification algorithm, which is then implemented
in C.
\autoref{sec:nfmt} also
compares its performance to that of drat-trim~\cite{heule-sat14}.
\autoref{sec:formal} then describes a formalization of the SAT problem in Coq,
which includes a specification of the pseudo-code in the previous section and a
proof of its soundness.
By applying the program extraction capabilities of Coq, we obtain a certified
checker in OCaml, which we evaluate on the same test set as our uncertified C
checker.
\autoref{sec:punchline} details the performance of the certified checker on the
verification of the proof of the Pythagorean Boolean Triples conjecture.
The paper concludes in~\autoref{sec:conc}.
\jnotef{
  Topics:
  \begin{enumerate}
  \item Importance of checking results of SAT solvers,
    e.g.~\cite{shankar-atva08}.
  \item Use of proof checking in SAT competitions~\cite{biere-jsat08,vangelder-isaim08}.
  \item Review other approaches for developing correct SAT solvers~\cite{smith-tr08,maric-tcs10,maric-lmcs11,weidenbach-ijcar16}.
  \item Brief review of SAT
    checking~\cite{goldberg-date03,malik-date03,biere-csr06,biere-sat06,biere-sat07,biere-jsat08,vangelder-isaim08,weber-jal09,dfms-coq09,dfms-ictac10,faure-cpp11,heule-cade13,heule-itp13,heule-fmcad13,heule-stvr14,heule-sat14,boudou-ijcar14,wetzler-phd15,heule-cade15,heule-sat16a}.
  \item Review optimizations to proof
    checking~\cite{lcf:psk:15a,lcf:psk:15b}.
  \end{enumerate}
}

\jnotef{
  We need to make sure we are not missing important references.\\
  For example, we need to look at the papers cited in the most recent
  papers of Heule et al.
}

\section{Preliminaries} \label{sec:prelim}

Standard Boolean Satisfiability (SAT) definitions are assumed
throughout~\cite{sat-handbook09}.
Propositional variables are taken from a set $X$.
In this work, we assume $X=\mathbb N^+$.
A literal is either
a variable or its negation. A clause is a disjunction of literals,
also viewed as a set of literals. A conjunctive normal form (CNF)
formula is a conjunction of clauses, also viewed as a set of clauses.
Formulas are represented in calligraphic font, e.g.~$\fml{F}$, with
$\var(\fml{F})$ denoting the subset of $X$ representing the variables
occurring in $\fml{F}$.
Clauses are represented with capital letters, e.g.~$C$.
Assignments are represented by a mapping $\mu:X\to\{0,1\}$, and the
semantics is defined inductively on the structure of propositional
formulas, as usual.
The paper focuses on CDCL SAT solvers~\cite{sat-handbook09}.
The symbol $\entails$ is used for entailment, whereas $\uprop$ is used
for representing the result of running the well-known unit propagation
algorithm.

This paper develops a formalized checker for proofs of
unsatisfiability of propositional formulas using the theorem prover
Coq~\cite{CoqArt}.
Coq is a type-theoretical constructive interactive theorem prover
based on the Calculus of Constructions (CoC)~\cite{Coq88} using a
propositions-as-types interpretation.
Proofs of theorems are terms in the CoC, which are constructed
interactively and type checked when the proof is completed; this
final step ensures that the correctness of the results obtained in Coq
only depends on the correctness of the type checker -- a short piece
of code that is much easier to verify by hand than the whole system.

A particular feature of Coq that we make use of in this paper is
program extraction~\cite{Letouzey2008}, which is an implementation of
the Curry--Howard correspondence for CoC and several functional
programming languages (in our case, OCaml).
Programs thus obtained are correct-by-construction, as they are
guaranteed to satisfy all the properties enforced by the Coq term they
originate form.
The CoC includes a special type \lstinline+Prop+ of propositions,
which are understood to have no computational content; in particular,
it is not allowed to define computational objects by case analysis on
a term whose type lives in \lstinline+Prop+.
This allows these terms to be removed by program extraction, making
the extracted code much smaller and more efficient; however, all
properties of the program that they express are still valid, as stated
by the soundness of the extraction mechanism.

\section{Propositional Proof Trace Formats} \label{sec:fmts}

\jnotef{
  Cover~\cite{malik-date03,goldberg-date03,biere-sat07,biere-jsat08,vangelder-isaim08,heule-cade13}.
}

The generation of resolution proof traces for checking the results of
SAT solvers has been actively studied since the early
2000s~\cite{goldberg-date03,malik-date03}.
Over the course of the years, different resolution proof tracing formats and
extensions have been
proposed~\cite{malik-date03,goldberg-date03,biere-csr06,biere-sat07,biere-jsat08,vangelder-isaim08,vangelder-amai12,heule-cade13,heule-itp13,heule-fmcad13,heule-stvr14,heule-sat14,heule-appa14,heule-fmcad14,heule-cade15,wetzler-phd15}.
These all boil down to listing information about the clauses learned
by CDCL SAT solvers, with recent efforts allowing an extended set of
operations~\cite{heule-cade13,heule-itp13}.
Resolution proof traces can list the literals of each learned
clause~\cite{vangelder-isaim08,vangelder-amai12,heule-fmcad13,heule-stvr14,heule-sat14,wetzler-phd15},
the labels of the clauses used for learning each clause, or
both~\cite{vangelder-isaim08,vangelder-amai12,biere-jsat08}.
Moreover, the checking of proof traces can traverse the trace from
the start to the end, i.e.\ \emph{forward checking}, or from end to
the start, i.e.\ \emph{backward checking}. In addition, the checking
of proof traces most often exploits one of two key mechanisms.
One validation mechanism uses trivial resolution steps
(TVR)~\cite{kautz-jair04}. This is a restriction over the already
restricted input resolution~\cite{vangelder-amai12}. For proof
checking purposes it suffices to require that every two consecutively
listed clauses \emph{must} contain a literal and its complement (and
obviously not be tautologous).
Another validation mechanism exploits the so-called reverse unit
propagation (RUP) property~\cite{goldberg-date03}.
Let $\fml{F}$ be a CNF formula, and $C$ be a clause learned from
$\fml{F}$. Thus, we must have $\fml{F}\entails C$.
The RUP property observes that, since $\fml{F}\land\neg
C\entails\bot$, then it is also true that $\fml{F}\land\neg
C\props\bot$.
The significance of the RUP property is that proof checking can be
reduced to validating a sequence of unit propagations that yield the
empty clause.
More recent work proposed RAT property
checking~\cite{heule-cade13,wetzler-phd15}. The resulting format,
DRAT, enables extended resolution proofs and, as a result, a wide
range of preprocessing
techniques~\cite{heule-cade13,heule-sat14,heule-cade15}.

A few additional properties of formats have important impact on the
type of resulting proof checking. Some formats do not allow for clause
deletion. This is the case with the
RUP~\cite{vangelder-isaim08,vangelder-amai12} and the
\emph{trace}~\cite{biere-jsat08} formats.
For formats that generate clause dependencies, some will allow clauses
\emph{not} to be ordered, and so the checker is required to infer the
correct order of steps.

\begin{example}
  \autoref{fig:fmts} samples the proof tracing formats RUP,
  \texttt{trace}, and DRUP. (Compared to DRUP, the DRAT format is of
  interest when extended resolution is used. Every DRUP proof is by definition also a DRAT proof.)
  With the exception of the more verbose RES format, earlier formats
  did not allow for clause deletion.
  The DRUP format (and the more recent DRAT format) allow for clause
  deletion. A number of different traces would represent DRAT traces,
  including the DRUP trace shown.
\end{example}
  
\begin{figure}[t]
  \begin{minipage}[t]{0.225\textwidth}
    \begin{alltt}
    problem CNF\\[0pt]
    p cnf 3 5
    \textcolor{darkgreen}{ 1  2 0}
    \textcolor{darkgreen}{-1  2 0}
    \textcolor{darkgreen}{ 1 -2 0}
    \textcolor{darkgreen}{-1  3 0}
    \textcolor{darkgreen}{-2 -3 0}
    \end{alltt}
  \end{minipage}
  \begin{minipage}[t]{0.2125\textwidth}
    \begin{alltt}
  RUP format\\[0pt]
  \textcolor{darkred}{1  0}
  \textcolor{darkred}{2  0}
  \textcolor{darkred}{3  0}
  \textcolor{darkred}{0}
    \end{alltt}
  \end{minipage}
  \begin{minipage}[t]{0.275\textwidth}
    \begin{alltt}
 tracecheck\\[0pt]
 1 \textcolor{darkgreen}{ 1  2 0} 0
 2 \textcolor{darkgreen}{-1  2 0} 0
 3 \textcolor{darkgreen}{ 1 -2 0} 0 
 4 \textcolor{darkgreen}{-1  3 0} 0
 5 \textcolor{darkgreen}{-2 -3 0} 0
 6  \textcolor{darkred}{1  0} \textcolor{darkyellow}{1 3} 0
 7  \textcolor{darkred}{2  0} \textcolor{darkyellow}{2 6} 0
 8  \textcolor{darkred}{3  0} \textcolor{darkyellow}{4 6} 0
 9  \textcolor{darkred}{0}  \textcolor{darkyellow}{5 7 8} 0
    \end{alltt}
    \end{minipage}
  \begin{minipage}[t]{0.225\textwidth}
    \begin{alltt}
DRUP format\\[0pt]
   \textcolor{darkred}{1  0}
d \textcolor{darkblue}{ 1  2 0}
d \textcolor{darkblue}{ 1 -2 0}
   \textcolor{darkred}{2  0}
d \textcolor{darkblue}{-1  2 0}
   \textcolor{darkred}{3  0}
d \textcolor{darkblue}{-1  3 0}
d \textcolor{darkblue}{ 1  0}
   \textcolor{darkred}{0}
    \end{alltt}
  \end{minipage}
  \caption{Examples of trace formats (example adapted
    from~\cite{heule-stvr14}; with original clauses in green, deletion information in blue, learnt clauses in red, and unit propagation information in yellow)}
  \label{fig:fmts}
\end{figure}

\autoref{tab:fmt-cmp} summarizes some of the best known formats, and
their drawbacks.
RES~\cite{vangelder-isaim08,vangelder-amai12} is extremely verbose,
separately encoding each resolution step, and is not in current use.
RUP~\cite{vangelder-isaim08,vangelder-amai12} and
\texttt{trace}~\cite{biere-jsat08} do not consider clause deletion,
and so are inadequate for modern SAT solvers.
DRUP addresses most of the drawbacks of earlier formats, and has been
superseded by DRAT, which provides an extended range of operations
besides clause learning.

\begin{table}[t]
  \begin{center}
    \caption{Comparison of some of the best known proof tracing formats}
    \label{tab:fmt-cmp}
    \hspace*{-0.35cm}
    \scalebox{0.75}{
      \renewcommand{\arraystretch}{1.25}
      \renewcommand{\tabcolsep}{0.45em}
      \begin{tabular}{|L{2.15cm}|C{2.25cm}|C{1.85cm}|C{1.85cm}|C{2cm}|C{1.85cm}|C{2cm}|}
        \hline
        Format     & Clause\newline Dependencies & Clause \newline Literals &
        Clause Deletion & Clause Reordering & RAT Checking & Drawbacks \\
        \hline\hline
        RES~\cite{vangelder-isaim08,vangelder-amai12} &
        Yes & Yes & Yes & No  & No  & Size, RAT \\ \hline
        RUP~\cite{vangelder-isaim08,vangelder-amai12} &
        No  & Yes & No  & No  & No  & Deletion, RAT \\ \hline
        \texttt{trace}~\cite{biere-jsat08} &
        Yes & Yes & No  & Yes & No  & Deletion, Reordering, RAT \\ \hline
        DRUP~\cite{heule-fmcad13,heule-stvr14} &
        No  & Yes & Yes & Yes & No  & RAT, Reordering \\ \hline
        DRAT~\cite{heule-sat14,wetzler-phd15} &
        No  & Yes & Yes & No & Yes & Complex checking  \\ \hline
      \end{tabular}
      \renewcommand{\arraystretch}{1.0}
      \renewcommand{\tabcolsep}{0.275em}
    }
  \end{center}
\end{table}
A number of guidelines for implementing resolution proof trace
checking have have emerged over the years.
First, backward checking is usually preferred, since only the clauses
in some unsatisfiable core need to be checked.
Second, RUP is preferred over checking TVR
steps~\cite{heule-cade13,heule-fmcad13,heule-sat14,heule-stvr14,wetzler-phd15},
because the format becomes more flexible.
Third, the SAT solver is often expected to minimize the time spent
generating the proof trace. This means that, for formats that output
clause dependencies, these are in general unordered. Moreover, modern
checkers also carry out the validation of the RAT
property~\cite{heule-cade13,heule-sat14,wetzler-phd15}.
These observations also indicate that recent work on checking of 
resolution proof traces has moved in the direction of more complex
checking procedures.

Besides efficient checking of resolution proof traces, another
important line of work has been to develop certified checkers. 
Different researchers exploited existing proof formats to develop
certified proof
checkers~\cite{weber-jal09,dfms-coq09,dfms-ictac10,faure-cpp11}.
The main drawback of this earlier work is that it was based on proof
formats that did not enable clause deletion. For large proofs, this
can result in unwieldy memory requirements.
Recent work addressed this issue by considering proof formats that
enable clause deletion~\cite{heule-cade13,heule-itp13,wetzler-phd15}.
Nevertheless, this recent work builds on complex proof checking
(see~\autoref{tab:fmt-cmp}) and so does not scale well in practice.

Given past evidence, one can argue that, in order to develop efficient
certified resolution proof checkers, proof checking \emph{must} be as
simple as possible. This has immediate consequences on the proof
format used, and also on the algorithm used for checking that format.
The next section details our proposed approach.
The proposed format requires enough information to enable a checking
algorithm that minimizes the processing effort. The actual checking
algorithms exploits the best features of TVR and RUP, to enable what
can be described as \emph{restricted reverse unit propagation}.

\section{Introducing the \nfmt Format} \label{sec:nfmt}
As described in the section above, an important aspect in the design of propositional proof trace formats
has been the desire to make it easy for SAT solvers to produce a proof in that format. As a
consequence, all the major proof trace formats have left some complex processing to the proof
checker:
\begin{itemize}
\item The DRUP and DRAT formats specify the clauses learnt, but they do
         not specify the clauses that are used in reverse unit propagation to verify redundancy of these clauses.
         Thus, proof checkers need to implement a full unit-propagation algorithm.
\item The \emph{trace} format specifies which clauses are used in reverse unit propagation, but
         it deliberately leaves the order of these undetermined. Thus, proof checkers
         still need to implement a unit-propagation algorithm, though limited to the clauses specified.
\end{itemize}
Experience from recent work verifying large-scale proof \cite{lcf:psk:15a,lcf:psk:15b}, co-authored by two of the authors of this work, suggests that fully eliminating complex processing is a key ingredient in developing efficient proof checkers that scale to very large proofs. Furthermore, in the concrete case of unit-propagation, efficient algorithms rely on pointer structures that are not easily ported to the typical functional programming setting used in most theorem provers.

Based on these observations, as well as on the importance of deleting clauses that are no longer needed~\cite{heule-cade13,heule-fmcad13}, we propose a novel proof trace format that includes deletion and fully eliminates complex processing, effectively reducing unit-propagation to simple pre-determined set operations.
\subsection{The Format}
The \emph{Generalized ResolutIon Trace (\nfmt)} format builds on the \emph{trace} format, but with two fundamental changes:
\begin{itemize}
\item We fix the order of the clauses dependencies given as a witness for each learnt clause to be one, in which unit propagation is able to produce the empty clause. This is a \emph{restriction} of the freedom allowed by the \emph{trace} format.
\item In addition to the two types of lines specifying original and learnt clauses, we \emph{extend} the format with a third type of line for deletions. These lines start with a \texttt{0} followed by a list of clause identifiers to delete and end with a \texttt{0}, and are thus easily distinguishable from the other two types of lines that start with a positive integer.
\end{itemize}
These changes are minimal w.r.t.~achieving the integration of deletion and the elimination of complex processing, and in particular the new lines keep some of the properties that make the \emph{trace} format easy to parse (two zeroes per line; the integers between those zeroes are clause identifiers). In this way, the changes follow the spirit of the extension of the RUP format to DRUP and later DRAT, just with \emph{trace} as the point of departure.

\autoref{fig:nfmt} shows how the \nfmt version of our running example from \autoref{fig:fmts} incorporates the deletion information from the DRUP format into a \emph{trace}-style proof, where the clause dependencies have been reordered
to avoid the complexity of checking the RUP property by full unit propagation, instead facilitating the application of restricted reverse unit propagation.
\begin{figure}[t]
  \begin{minipage}[t]{0.225\textwidth}
    \begin{alltt}
    problem CNF\\[0pt]
    p cnf 3 5
    \textcolor{darkgreen}{ 1  2 0}
    \textcolor{darkgreen}{-1  2 0}
    \textcolor{darkgreen}{ 1 -2 0}
    \textcolor{darkgreen}{-1  3 0}
    \textcolor{darkgreen}{-2 -3 0}
    \end{alltt}
  \end{minipage}
  \begin{minipage}[t]{0.275\textwidth}
    \begin{alltt}
  tracecheck\\[0pt]
  1 \textcolor{darkgreen}{ 1  2 0} 0
  2 \textcolor{darkgreen}{-1  2 0} 0
  3 \textcolor{darkgreen}{ 1 -2 0} 0
  4 \textcolor{darkgreen}{-1  3 0} 0
  5 \textcolor{darkgreen}{-2 -3 0} 0
  6  \textcolor{darkred}{1  0} \textcolor{darkyellow}{1 3} 0
  7  \textcolor{darkred}{2  0} \textcolor{darkyellow}{2 6} 0
  8  \textcolor{darkred}{3  0} \textcolor{darkyellow}{4 6} 0
  9  \textcolor{darkred}{0}  \textcolor{darkyellow}{5 7 8} 0
    \end{alltt}
    \end{minipage}
  \begin{minipage}[t]{0.225\textwidth}
    \begin{alltt}
 DRUP format\\[0pt]
    \textcolor{darkred}{1  0}
 d \textcolor{darkblue}{ 1  2 0}
 d \textcolor{darkblue}{ 1 -2 0}
    \textcolor{darkred}{2  0}
 d \textcolor{darkblue}{-1  2 0}
    \textcolor{darkred}{3  0}
 d \textcolor{darkblue}{-1  3 0}
 d \textcolor{darkblue}{ 1  0}
    \textcolor{darkred}{0}
    \end{alltt}
  \end{minipage}
  \begin{minipage}[t]{0.2125\textwidth}
    \begin{alltt}
\nfmt format\\[0pt]
1 \textcolor{darkgreen}{ 1  2 0} 0
2 \textcolor{darkgreen}{-1  2 0} 0
3 \textcolor{darkgreen}{ 1 -2 0} 0
4 \textcolor{darkgreen}{-1  3 0} 0
5 \textcolor{darkgreen}{-2 -3 0} 0
6  \textcolor{darkred}{1  0} \textcolor{darkyellow}{1 3} 0
0  \textcolor{darkblue}{1  3} 0
7  \textcolor{darkred}{2  0} \textcolor{darkyellow}{6 2} 0
0  \textcolor{darkblue}{2}  0
8  \textcolor{darkred}{3  0} \textcolor{darkyellow}{6 4} 0
0  \textcolor{darkblue}{4  6} 0
9  \textcolor{darkred}{0}  \textcolor{darkyellow}{7 8 5} 0
    \end{alltt}
  \end{minipage}
  \caption{Synthesis of the \nfmt format (with original clauses in green, deletion information in blue, learnt clauses in red, and unit propagation information in yellow).}
  \label{fig:nfmt}
\end{figure}

The full syntax of the \nfmt format is given by the grammar in \autoref{fig:syntax}, where for the sake of sanity whitespace (tabs and spaces) is ignored.
\begin{figure}[t]
\begin{center}
\newcommand{\sNon}[1]{\ensuremath{\langle#1\rangle}}
\newcommand{\sTerm}[1]{\ensuremath{"#1"}}
\newcommand{\sStar}[1]{\ensuremath{\{ #1 \}}}
\newcommand{\sAlt}[2]{\ensuremath{ #1 \mid #2 }}
\newcommand{\sConc}[2]{\ensuremath{#1, #2}}
\newcommand{\sPar}[1]{\ensuremath{(#1)}}
\newcommand{\sProof}{\sNon{proof}}
\newcommand{\sLine}{\sNon{line}}
\newcommand{\sOriginal}{\sNon{original}}
\newcommand{\sLearnt}{\sNon{learnt}}
\newcommand{\sDelete}{\sNon{delete}}
\newcommand{\sId}{\sNon{id}}
\newcommand{\sIdlist}{\sNon{idlist}}
\newcommand{\sClause}{\sNon{clause}}
\newcommand{\sPos}{\sNon{pos}}
\newcommand{\sNeg}{\sNon{neg}}
\newcommand{\sLit}{\sNon{lit}}
\begin{tabular}{lclclcl}
\sProof & = & \sStar{\sLine}
&\qquad
&\sClause & = & \sConc{\sStar{\sLit}}{\sTerm{0}}\\
\sLine & = & \sConc{\sPar{\sAlt{\sOriginal}{\sAlt{\sLearnt}{\textcolor{darkgreen}{\sDelete}}}}}{\sTerm{\backslash n}}
&&\sIdlist & = & \sConc{\sId}{\sStar{\sId}}\\
\sOriginal & = & \sConc{\sId}{\sConc{\sClause}{\sConc{\sTerm{0}}{\sTerm{0}}}}
&&\sId & = & \sPos\\
\sLearnt & = & \sConc{\sId}{\sConc{\sClause}{\sConc{\sTerm{0}}{\sConc{\textcolor{darkred}{\sIdlist}}{\sTerm{0}}}}}
&&\sLit & = & \sAlt{\sPos}{\sNeg}\\
\textcolor{darkgreen}{\sDelete} & \textcolor{darkgreen}{=} & \textcolor{darkgreen}{\sConc{\sTerm{0}}{\sConc{\sIdlist}{\sTerm{0}}}}
&&\sPos & = & \sAlt{\sTerm{1}}{\sAlt{\sTerm{2}}{\ldots}}\\
&&&&\sNeg & = & \sConc{\sTerm{\text{-}}}{\sPos}\\
\end{tabular}
\end{center}
  \caption{EBNF grammar for the \nfmt format.}
  \label{fig:syntax}
\end{figure}
Here, additions with respect to the original \emph{trace} format are given in green. In addition to the extension
with delete information, there is a semantic restriction on the list of clause identifiers marked in red, namely that the
clause dependencies represented are in the order as specified above.
Existing parsers for the \emph{trace} format should be easy to extend to this syntax.

\subsection{The Checker}

To obtain an empirical evaluation of the potential of the \nfmt format, we implemented a
proof checking algorithm based on restricted reverse unit propagation in C. The source code is available from \cite{grit}. While the C code is quite optimized, the general algorithm follows the pseudo code given in \autoref{fig:py-checker} as $25$ lines of fully-functional Python (also available from \cite{grit}).
\begin{figure}[t]
\begin{center}
\begin{lstlisting}
def parse(line):
  ints = [int(s) for s in line.split()]
  i0 = ints.index(0)
  return ints[0], set(ints[1:i0]), ints[i0+1:-1]
def verify(file):
  cs = {}
  for id, c, ids in (parse(line) for line in file):
    if not id:     # delete clauses
      for id in ids:  del cs[id]
    elif not ids:  # add original clause
      cs[id] = c  
    else:          # check & add learnt clause
      d = c.copy()
      for i in ids:
        e = cs[i]-d
        if e:
          d.add(-e.pop())  # propagate
          assert not e     # is unit?
        else:  # empty clause reached
          cs[id] = c       
          if not c:  return "VERIFIED"
          break
  return "NOT VERIFIED"
import sys
print(verify(open(sys.argv[1])))
\end{lstlisting}
\end{center}
  \caption{Fully functional checker for the \nfmt format written in Python.}
  \label{fig:py-checker}
\end{figure}

The set of instances we considered consists of the $280$ instances from the 2015 and 2016 main and parallel tracks of the SAT competition that could be shown to be UNSAT within $5000$ seconds using the 2016 competition version of lingeling. For each of these instances, the original CNF and proof trace are trimmed and optimized using drat-trim in backward checking mode. This is a side-effect of using drat-trim to generate proof traces in the \nfmt format, and was applied in the same way to generate DRAT files from the original RUP files in order to ensure a level playing field.

\begin{figure}[t]
\begin{center}
\includegraphics[scale=0.4]{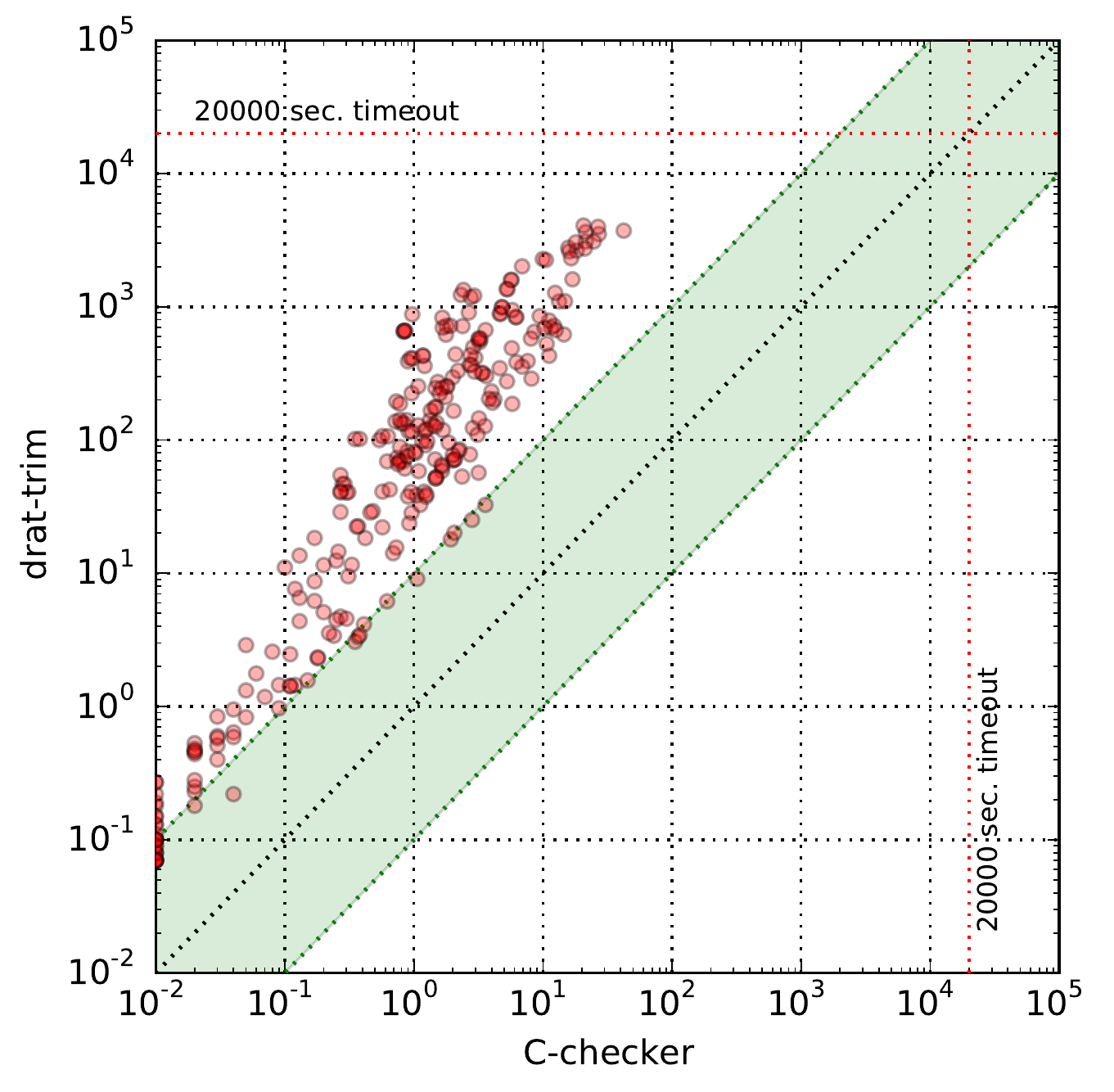}
\includegraphics[scale=0.4]{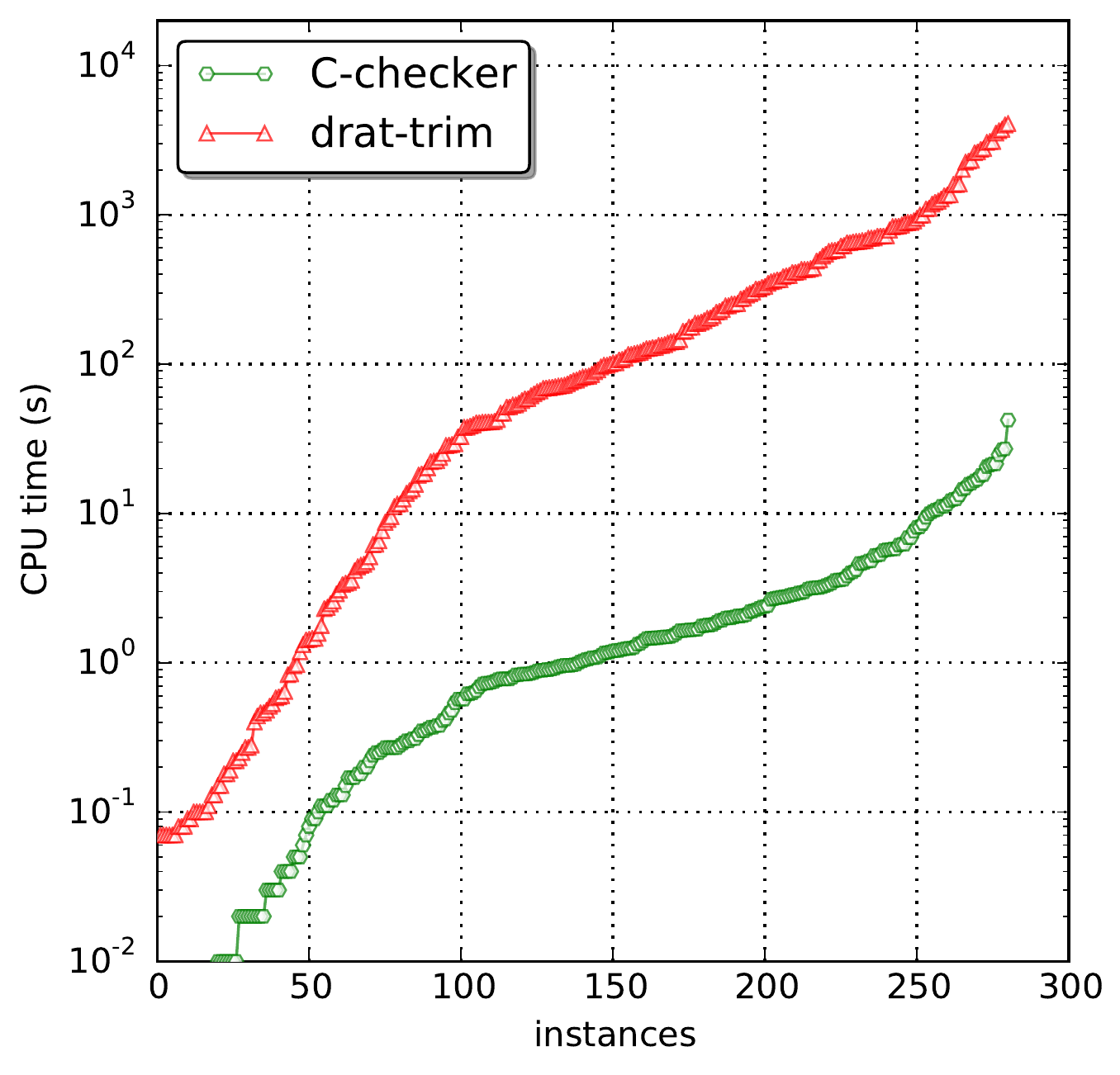}
\end{center}
  \caption{Scatter and cactus plot comparing the runtime of the C-checker on \nfmt files and drat-trim on the corresponding DRAT files.}
  \label{fig:dtvsc-scatter}
\end{figure}

The C-checker successfully verifies all $280$ \nfmt files in just over $14$ minutes ($843.64$ s), while drat-trim requires more than a day to solve the corresponding DRAT files ($109214.08$ s) using backward mode. Executing drat-trim in forward mode incurred a runtime overhead of $15$\% on the total set of trimmed and optimized instances. As expected, the overhead was even bigger when working on the original CNFs and proof traces. The quantitative results are summarized in the plots of \autoref{fig:dtvsc-scatter}, with details available from \cite{grit}.

This two-orders-of-magnitude speedup demonstrates the potential of using a file format for propositional resolution proof checking by restricted reverse unit propagation. Note that we currently do \emph{not} output the \nfmt format directly, but require a modified version of drat-trim as a pre-processor\footnote{The modified version essentially uses drat-trim's tracecheck output, interleaving it with deletion information. The modified source code is available from \cite{grit}.} in order to determine both the order of clauses used in unit propagation, the set of original and learnt clauses relevant, and the deletion of clauses that are no longer needed.
We stress the importance of this additional information in obtaining the performance gains we measure.
Additional experiments (whose results we do not detail for space constraints) show that deletion of clauses alone is responsible for a speedup of more than one order of magnitude for the larger instances, when using the certified checker we develop in the next section.

While it is in principle thinkable to modify a SAT solver to output the \nfmt format directly, building on \cite{vangelder-sat09}, in this work our focus is on enabling sufficiently efficient certified proof checking. To this end, it seems fully acceptable to run an uncertified proof checker as a pre-processor to generate the oracle data enabling the application of restricted reverse unit propagation in a certified checker.

\section{Coq Formalization}
\label{sec:formal}

We now describe a Coq formalization that yields a certified checker of SAT
proofs.
We follow the strategy outlined in~\cite{lcf:psk:15a,lcf:psk:15b}: first, we
formalize the necessary theoretical concepts (propositional satisfiability,
entailment and soundness of unit propagation); then, we naively specify the verification
algorithm; finally, we optimize this algorithm using standard computer science
techniques to obtain feasible runtimes.
In the interest of succintness, we only present the formalization obtained at
the end of this process.
The source files can be obtained from~\cite{grit}.

\subsection{Formalizing propositional satisfiability}

We identify propositional variables with Coq's binary natural numbers (type \lstinline+positive+), and
define a literal to be a signed variable.
The type of literals is thus isomorphic to that of integers (excluding zero).

\begin{lstlisting}
Inductive Literal : Type :=
  | pos : positive -> Literal
  | neg : positive -> Literal.
\end{lstlisting}

A clause is a set of literals, and a CNF is a set of clauses.
For efficiency, there are two different definitions of each type, with mappings
between them.
A \lstinline+Clause+ is a \lstinline+list Literal+, and is the type preferably
used in proofs due to its simplicity; it is also the type used for inputting
data from the oracle.
A \lstinline+CNF+ is a \lstinline+BinaryTree Clause+, where the dependent type
\lstinline+BinaryTree+ implements search trees over any type with a comparison
operator.
This is the type of the CNF given as input to the algorithm, which is built once,
never changed, and repeatedly tested for membership.
The working set uses two different representations of these types.
A \lstinline+SetClause+ is a \lstinline+BinaryTree Literal+, where in particular
set differences can be computed much more efficiently than using
\lstinline+Clause+.
Finally, an \lstinline+ICNF+ is a \lstinline+Map {cl:SetClause | SC_wf cl}+,
where \lstinline+Map+ is the Coq standard library's implementation of Patricia
trees.
The elements of an \lstinline+ICNF+ must be well-formed search trees (ensured by
the condition in the definition of subset type); proofs of well-formedness
do not contain computational meaning and are removed by extraction.
In particular, every \lstinline+SetClause+ built from a \lstinline+Clause+ is
well-formed.

A valuation is a function from positive numbers to Booleans.
Satisfaction is defined for literals, clauses and CNFs either directly (as
below) or by translating to the appropriate type (for \lstinline+SetClause+ and
\lstinline+ICNF+).

\begin{lstlisting}
Definition Valuation := positive -> bool.

Fixpoint L_satisfies (v:Valuation) (l:Literal) : Prop :=
  match l with
  | pos x => if (v x) then True else False
  | neg x => if (v x) then False else True
  end.

Fixpoint C_satisfies (v:Valuation) (c:Clause) : Prop :=
  match c with
  | nil => False
  | l :: c' => (L_satisfies v l) \/ (C_satisfies v c')
  end.

Fixpoint satisfies (v:Valuation) (c:CNF) : Prop :=
  match c with
  | nought => True
  | node cl c' c'' => (C_satisfies v cl) /\ (satisfies v c') /\ (satisfies v c'')
  end.

Definition unsat (c:CNF) : Prop := forall v:Valuation, ~(satisfies v c).

Definition entails (c:CNF) (c':Clause) : Prop :=
  forall v:Valuation, satisfies v c -> C_satisfies v c'.
\end{lstlisting}

We then prove the intuitive semantics of satisfaction: a clause is satisfied
if one of its literals is satisfied, and a CNF is satisfied if all its clauses
are satisfied.
Other properties that we need include: the empty clause is unsatisfiable; every
non-empty clause is satisfiable; a subset of a satisfiable CNF is satisfiable;
and a CNF that entails the empty clause is unsatisfiable.

\begin{lstlisting}
Lemma C_satisfies_exist : forall (v:Valuation) (cl:Clause),
  C_satisfies v cl -> exists l, In l cl /\ L_satisfies v l.

Lemma satisfies_remove : forall (c:CNF) (cl:Clause) (v:Valuation),
  satisfies v c -> satisfies v (CNF_remove cl c).

Lemma unsat_subset : forall (c c':CNF),
  (forall cl, CNF_in cl c -> CNF_in cl c') -> unsat c -> unsat c'.

Lemma CNF_empty : forall c, entails c nil -> unsat c.
\end{lstlisting}

\subsection{Soundness of unit propagation}

The key ingredient to verifying unsatisfiability proofs in \nfmt format
is being able to verify the original unit propagation steps.
Soundness of unit propagation relies on the following results, formalizing the two relevant
outcomes of resolving two clauses: a unit clause and the empty clause.

\begin{lstlisting}
Lemma propagate_singleton : forall (cs:CNF) (c c':SetClause), forall l,
  entails cs (SetClause_to_Clause (SC_add (negate l) c')) ->
  SC_diff c c' = (node l nought nought) -> entails (CNF_add c cs) c'.

Lemma propagate_empty : forall (cs:CNF) (c c':SetClause),
  SC_diff c c' = nought -> entails (BT_add Clause_compare c cs) c'.
\end{lstlisting}

We then define the propagation step: given an \lstinline+ICNF+, a
\lstinline+SetClause+ and a list of indices (of type \lstinline+ad+, used in the
implementation of \lstinline+Map+), we successively check whether the set
difference between the given clause and the clause with the first index from the
list is empty\footnote{When \lstinline+c+ is a clause, \lstinline+(SetClause_eq_nil_cons c)+ checks whether
  \lstinline+c+ is the empty clause, otherwise returning one of its literals and a clause with the remaining ones.}
 (in which case we return \lstinline+true+), a singleton (in which
case we add the negation of the derived literal to the clause, remove the index
from the list and recur), or a longer list of literals (and we return
\lstinline+false+).
If the result is true, the initial clause is entailed by the original
\lstinline+ICNF+.

\begin{lstlisting}
Fixpoint propagate (cs:ICNF) (c:SetClause) (is:list ad) : bool := 
  match is with
  | nil => false
  | (i::is) => match SetClause_eq_nil_cons (SC_diff (get_ICNF cs i) c) with
    | inright _ => true
    | inleft H => let (l,Hl) := H in let (c',Hc) := Hl in
        match SetClause_eq_nil_cons c' with
        | inleft _ => false
        | inright _ => let (c'',_) := Hc in
            match SetClause_eq_nil_cons c'' with
            | inleft _ => false
            | inright _ => propagate cs (SC_add (negate l) c) is
  end end end end.

Lemma propagate_sound : forall (cs:ICNF) (c:SetClause) (is:list ad),
  propagate cs c is = true -> entails cs c.
\end{lstlisting}

Observe that \lstinline+propagate+ is implementing a restricted
version of reverse unit propagation, which in particular avoids
complex processing for the next clause to use in unit propagation.

To check that a given formula is unsatisfiable, we start with an empty working
set, and iteratively change it by applying actions given by the oracle.
These can have three types: delete a clause; add a clause from the original CNF;
or extend it with a clause that is derivable by unit propagation (and the oracle
provides the clauses that should be used in this derivation).

\begin{lstlisting}
Inductive Action : Type :=
  | D : list ad -> Action
  | O : ad -> Clause -> Action
  | R : ad -> Clause -> list ad -> Action.

Definition Answer := bool.

Function refute_work (w:ICNF) (c:CNF) (Hc:CNF_wf c) (O:Oracle)
  {measure Oracle_size O} : Answer :=
  match (force O) with
  | lnil => false
  | lcons (D nil) O' => refute_work w c Hc O'
  | lcons (D (i::is)) O' => refute_work (del_ICNF i w) c Hc (lazy (lcons (D is) O'))
  | lcons (O i cl) O' => if (BT_in_dec _ _ _ __ cl c Hc)
                         then (refute_work (add_ICNF i cl' _ w) c Hc O') else false
  | lcons (R i nil is) O' => propagate w nought is
  | lcons (R i cl is) O' => andb (propagate w cl' is)
                                 (refute_work (add_ICNF i cl' _ w) c Hc O')
  end.

Definition refute (c:list Clause) (O:Oracle) : Answer :=
  refute_work empty_ICNF (make_CNF c) (make_CNF_wf c) O.
\end{lstlisting}

Function \lstinline+refute_work+ implements this process over an oracle that is
defined to be a lazy list.
Lazy lists are defined exactly as lists, but with the second argument of
\lstinline+lcons+ inside an invocation of an identity function.
Likewise, \lstinline+lazy+ and \lstinline+force+ are defined as the
identity.
These additions are necessary to be able to extract a memory-efficient checker
to OCaml. On extraction, these functions are mapped to the adequate OCaml
constructs implementing laziness; although in principle this approach could break
soundness of extraction, these constructs do indeed behave as identities.
Without them, the extracted checker attempts to load the entire oracle data
at the start of execution, and thus risks running out of memory for larger proofs.%
\footnote{Targeting a lazy language like Haskell would not require this workaround. However, in our context, using OCaml reduced computation times to around one-fourth.}

The initialization of auxiliary variables is done in \lstinline+refute+, which
only receives a CNF (in list form) and the oracle.
This function then calls \lstinline+refute_work+ with an empty working set and
the clause in binary tree representation.
For legibility, we replaced proof obligations that are removed by extraction to
an underscore, and wrote \lstinline+cl'+ for the result of converting
\lstinline+cl+ to a \lstinline+SetClause+.

The following result states soundness of \lstinline+refute+: if
\lstinline+refute c O+ returns \lstinline+true+, then \lstinline+c+ is unsatisfiable.
Since \lstinline+O+ is universally quantified, the result holds even if the
oracle gives incorrect data.
(Namely, because \lstinline+refute+ will output \lstinline+false+.)

\begin{lstlisting}
Theorem refute_correct : forall c O, refute c O = true -> unsat (make_CNF c).
\end{lstlisting}

\subsection{Experimental Evaluation}
In order to evaluate the efficiency of our formalized checker, we extracted it
to OCaml. The extraction definition is available in the file \lstinline+Extraction.v+ from \cite{grit}.
As is customary, we extract the Coq type \lstinline+positive+, used for variable
and clause identifiers, to OCaml's native integers, and the comparator function
on this type to a straightforward implementation of comparison of two integers.
This reduces not only the memory footprint of the verified checker, but also its
runtime (as lookups in \lstinline+ICNF+s require comparison of keys).
It is routine to check that these functions are correct.
Furthermore, as described above, we extract the type of lazy lists to OCaml's
lazy lists.

We ran the certified extracted checker on the same $280$ unsatisfiable instances
as in the previous section, with a timeout of $20{,}000$ seconds, resulting in
$260$ successful verifications and $20$ timeouts.
On the $260$ examples, the certified checker runs in good $4$ days and $18$
hours ($412469.50$s) compared to good $2$ days and $17$ hours ($234922.46$s)
required by the uncertified checker drat-trim.
The pre-processing using our modified version of drat-trim adds another $2$ days
and $19$ hours ($241453.84$s) for a total runtime of $7$ days and good $13$ ($653923.34$s).
Thus, the extra degree of confidence provided by the certified checker comes at
the price of approx.~$2.8$ times slower verification for these instances ($180$\% overhead).

\begin{figure}[t]
\begin{center}
\includegraphics[scale=0.4]{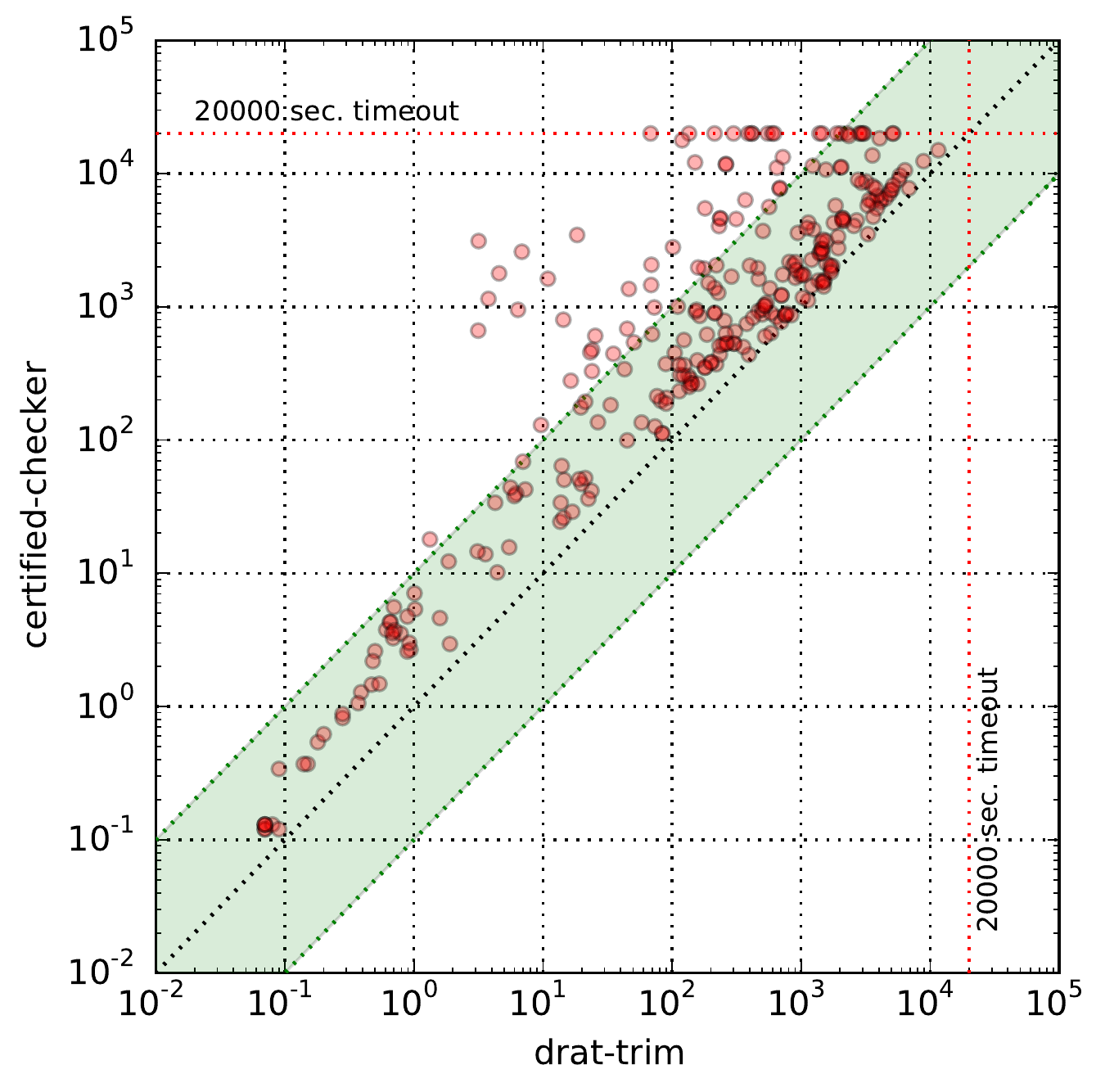}
\includegraphics[scale=0.4]{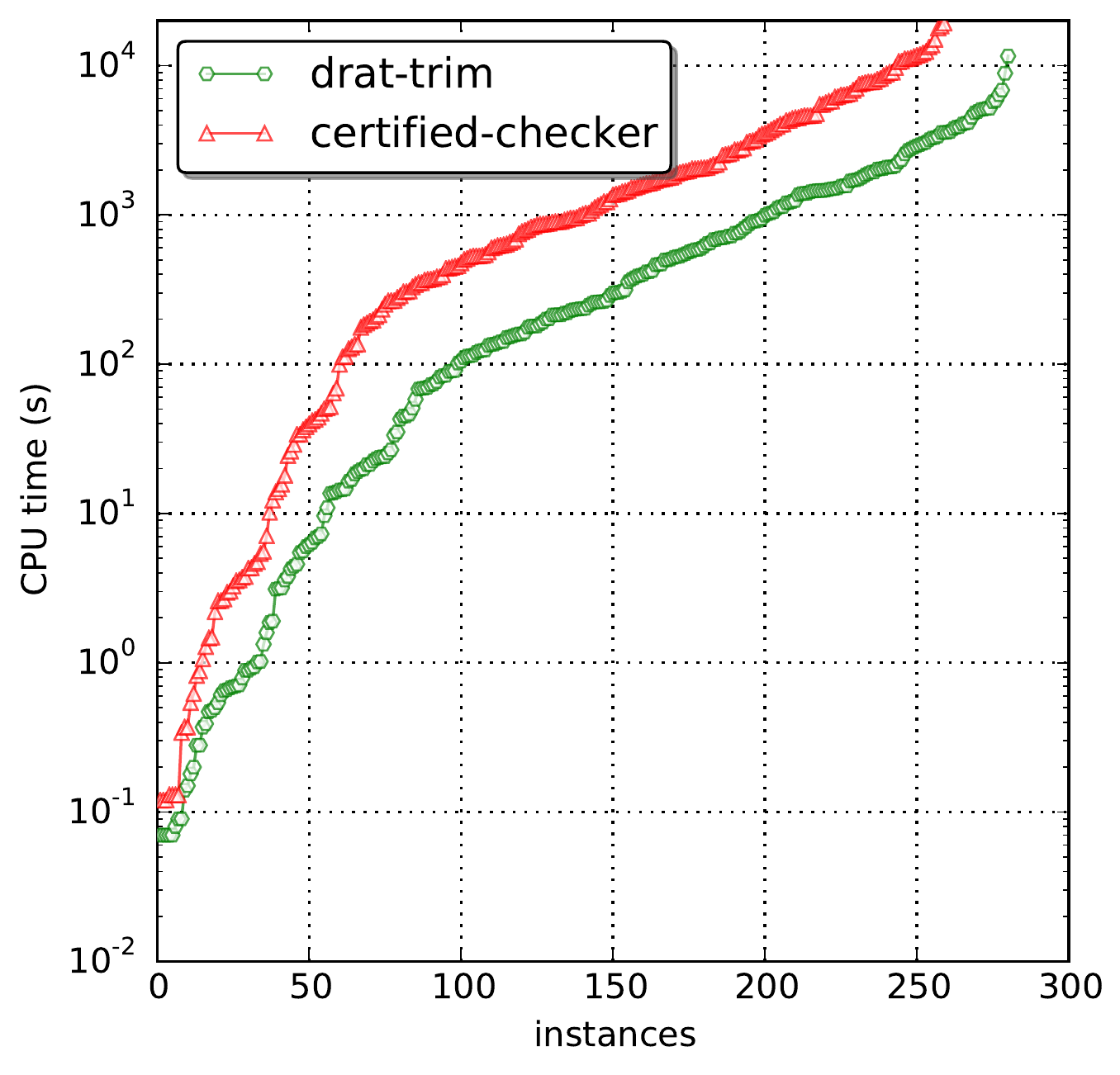}
\end{center}
  \caption{Scatter and cactus plot comparing the runtime of our certified checker (including pre-processing) and drat-trim on the original proof traces from lingeling.}
  \label{fig:dtvscoq-scatter}
\end{figure}

The quantitative results on all $280$ instances are summarized in the plots of \autoref{fig:dtvscoq-scatter}, where
we added the pre-processing time to the time of the certified checker, with details available from~\cite{grit}.

\begin{figure}[t]
\begin{center}
\includegraphics[scale=0.4]{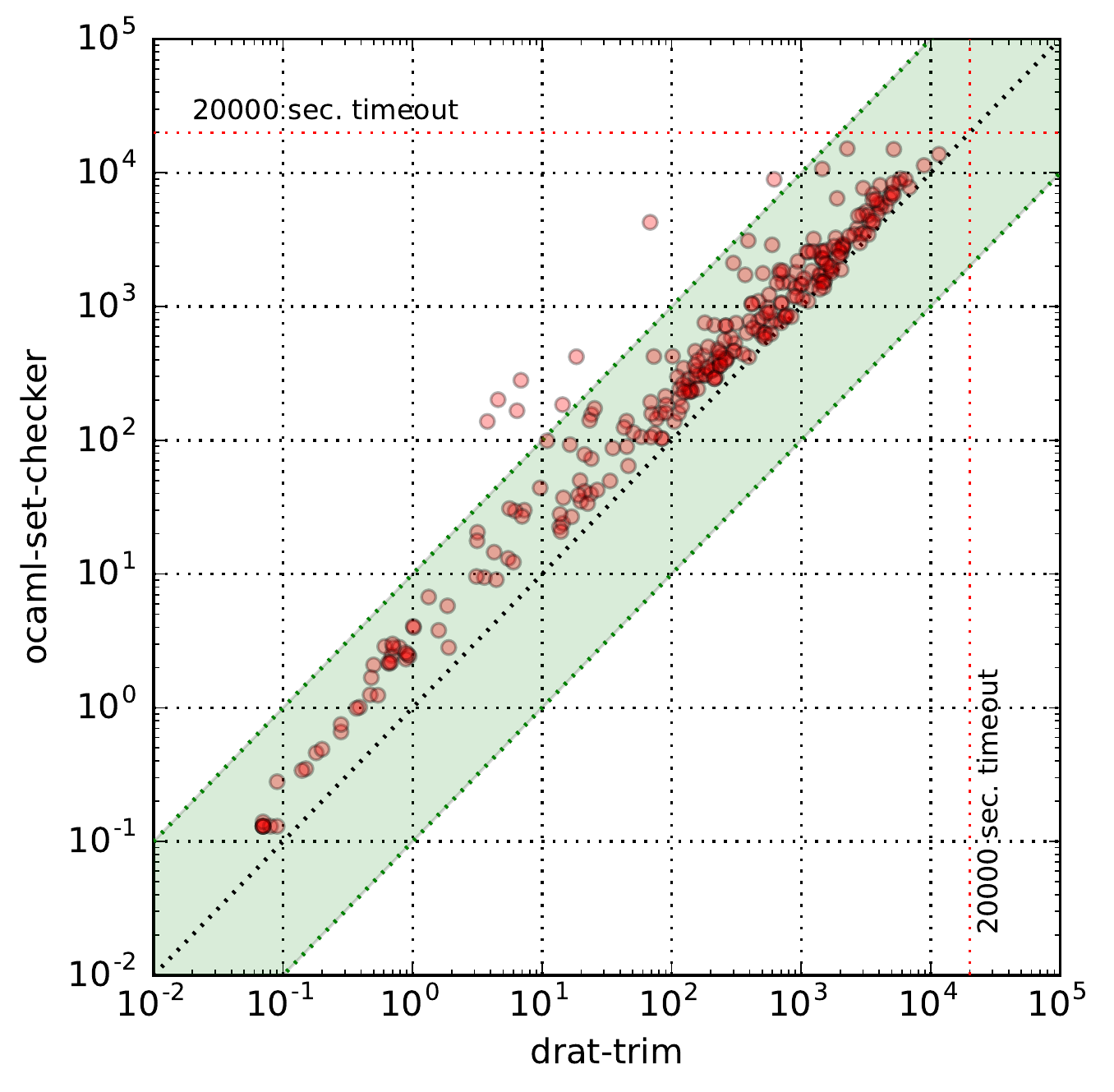}
\includegraphics[scale=0.4]{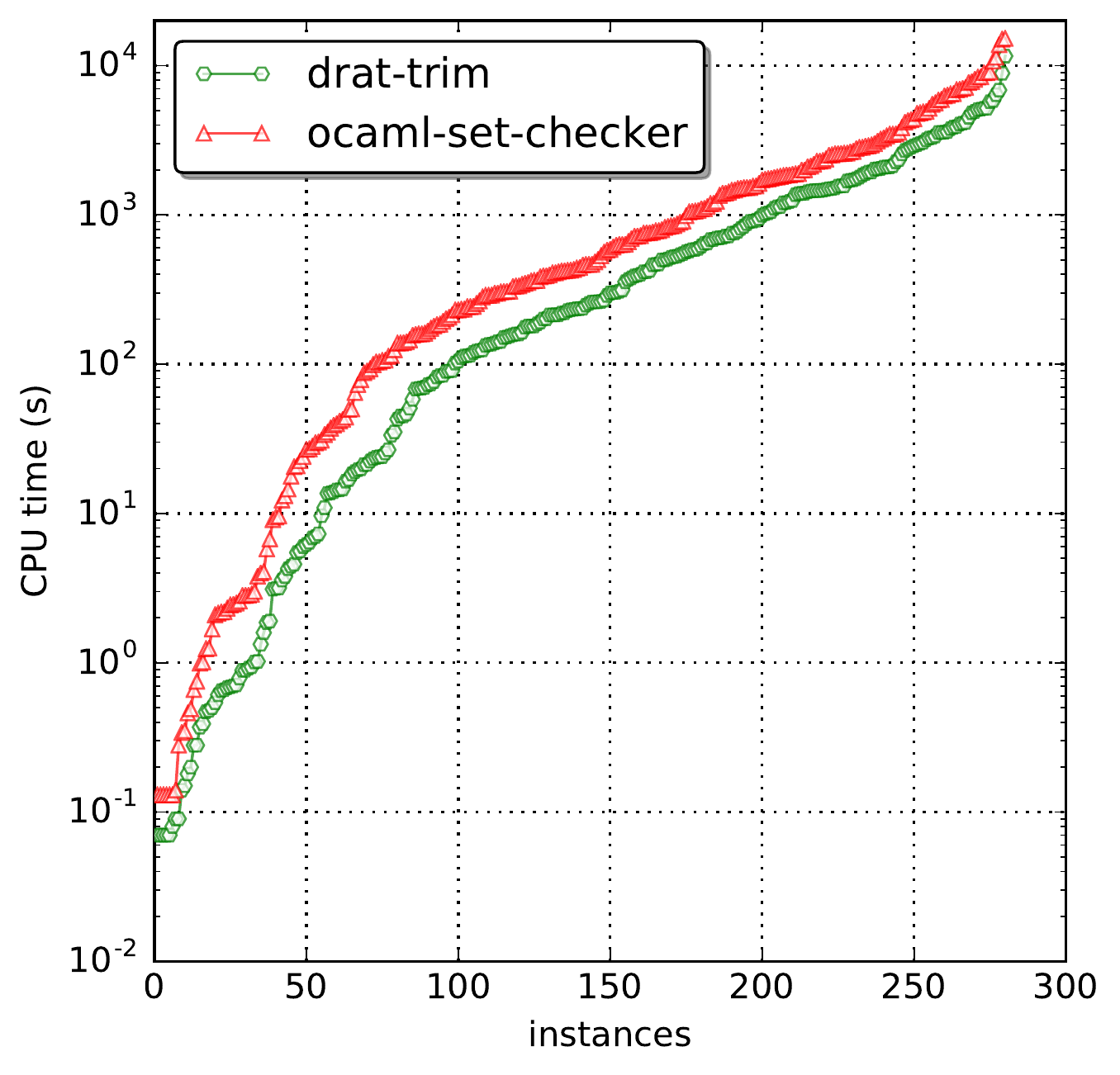}
\end{center}
  \caption{Scatter and cactus plot comparing the runtime of a certified checker using OCaml sets (including pre-processing) and drat-trim on the original proof traces from lingeling.}
  \label{fig:dtvsnative-scatter}
\end{figure}

The reason for the $20$ timeouts can be found in the set implementation of our formalization. If we extract Coq sets to native OCaml sets, there are no time-outs. We extracted such a version of the certified checker in order to check this hypothesis, as well as to assess the performance impact. And indeed, this version of our checker successfully verifies all $280$ \nfmt files in less time ($186599.20$s) than it takes to pre-process them using our modified drat-trim version ($281516.13$), and consequently the overhead of running a certified checker instead of an uncertified checker is down to $75$\%. The quantitative results for this variant are summarized in the plots of \autoref{fig:dtvsnative-scatter}, with details available from~\cite{grit}.

\section{Veryifing the Boolean Pythagorean Triples proof}
\label{sec:punchline}
As a large-scale litmus test of our formally verified checker, we reconstituted the recent SAT-based proof of the Boolean Pythagorean Triples conjecture \cite{heule-sat16a} ($508$ CPU days) using the incremental SAT solver \emph{iGlucose}, transformed it into the \nfmt format ($871$ CPU days) using our modified version of drat-trim, and formally verified that all $1{,}000{,}000$ cases (``cubes'') cover the entire search space ($12$ minutes), and that they are all indeed unsatisfiable ($2608$ days) using our certified checker (the original version, where all data structures except integers are extracted).
This amounts to formally verifying the Boolean Pythagorean Triples conjecture (provided that its encoding as a propositional formula is correct).
\begin{figure}[t]
\begin{center}
\includegraphics[scale=0.55]{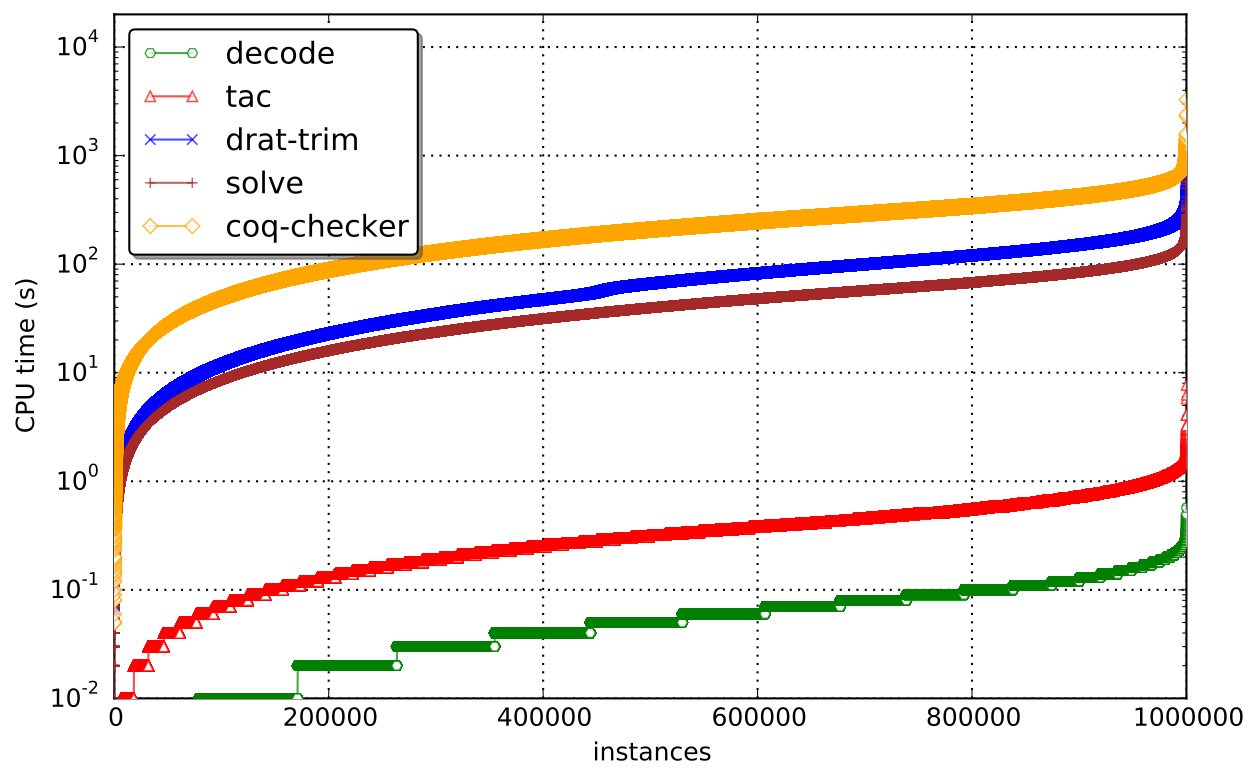}
\end{center}
  \caption{Cactus plot comparing the runtimes for reconstituting the proof (decode and solve), transforming it into \nfmt (drat-trim and tac), and formally verifying the \nfmt files using our certified checker.}
  \label{fig:user-cactus}
\end{figure}

The cactus plot in \autoref{fig:user-cactus} visualizes the distribution of runtime on the $1{,}000{,}000$ cubes.
The size of the reconstituted proof traces in RUP format was measured to be $175$ TB. After transformation to the more detailed \nfmt format, the proof traces filled a total $389$ TB. During runtime, the maximum resident memory usage of the incremental SAT solver was $237$ MB, while drat-trim in backward mode used up to $1.59$ GB. Our certified checker reached a maximum of $67$ MB of resident memory usage thanks to lazyness.

\section{Conclusions \& Research Directions} \label{sec:conc}

This paper revisits past work on proof checking, aiming at developing
high-performance certified proof checkers. It
proposes a new format, which enables a very simple proof checking
algorithm. This simple algorithm is formalized in the Coq theorem
prover, from which an OCaml executable is then extracted.

The experimental results amply demonstrate the validity of the
proposed approach. The C implementation of the checker is on average
two orders of magnitude faster than what can be considered a reference
C-implemented proof checker, drat-trim~\cite{heule-sat14,drat-trim}.
This represents an essential requirement for developing an efficient
certified proof checker.
More importantly, the certified OCaml version of the checker performs
comparably with drat-trim on problem instances from the SAT
competitions. Perhaps more significantly, the certified checker has
been used to formally verify the $200$ TB proof of the Boolean Pythagorean
Triples conjecture~\cite{heule-sat16a}, in time comparable to the
non-certified drat-trim checker.

Future work will address existing limitations of the approach.
Currently, a modified version of drat-trim is used to generate the
\nfmt format. This can impact the overall running time, especially if
the C-implemented checker for the \nfmt format is to be used.
This also includes modifying top performing SAT solvers to output the
\nfmt format, potentially based on A.~Van Gelder's
approach~\cite{vangelder-sat09,vangelder-amai12}.
In addition, although not the focus on this paper, a natural extension
of this work is to extend \nfmt to be as general a format as DRAT, in
particular by including support for the RAT property.

\paragraph{Acknowledgements.} 
The authors thank M.~Heule for comments on an early draft of this
paper.

\clearpage
\bibliographystyle{abbrv}
\bibliography{paper}

\begin{thebibliography}{10}

\bibitem{mehlhorn-jar14}
E.~Alkassar, S.~B{\"{o}}hme, K.~Mehlhorn, and C.~Rizkallah.
\newblock A framework for the verification of certifying computations.
\newblock {\em J. Autom. Reasoning}, 52(3):241--273, 2014.

\bibitem{faure-cpp11}
M.~Armand, G.~Faure, B.~Gr{\'{e}}goire, C.~Keller, L.~Th{\'{e}}ry, and
  B.~Werner.
\newblock A modular integration of {SAT/SMT} solvers to {Coq} through proof
  witnesses.
\newblock In {\em CPP}, pages 135--150, 2011.

\bibitem{kautz-jair04}
P.~Beame, H.~A. Kautz, and A.~Sabharwal.
\newblock Towards understanding and harnessing the potential of clause
  learning.
\newblock {\em J. Artif. Intell. Res. {(JAIR)}}, 22:319--351, 2004.

\bibitem{CoqArt}
Y.~Bertot and P.~Cast{\'e}ran.
\newblock {\em Interactive Theorem Proving and Program Development}.
\newblock Texts in Theoretical Computer Science. Springer, 2004.

\bibitem{biere-jsat08}
A.~Biere.
\newblock {PicoSAT} essentials.
\newblock {\em {JSAT}}, 4(2-4):75--97, 2008.

\bibitem{sat-handbook09}
A.~Biere, M.~Heule, H.~van Maaren, and T.~Walsh, editors.
\newblock {\em Handbook of Satisfiability}, volume 185 of {\em Frontiers in
  Artificial Intelligence and Applications}. {IOS} Press, 2009.

\bibitem{weidenbach-ijcar16}
J.~C. Blanchette, M.~Fleury, and C.~Weidenbach.
\newblock A verified {SAT} solver framework with learn, forget, restart, and
  incrementality.
\newblock In {\em IJCAR}, pages 25--44, 2016.

\bibitem{blum-stoc89}
M.~Blum and S.~Kannan.
\newblock Designing programs that check their work.
\newblock In {\em STOC}, pages 86--97, 1989.

\bibitem{selman-cp14}
R.~L. Bras, C.~P. Gomes, and B.~Selman.
\newblock On the {E}rd{\H{o}}s discrepancy problem.
\newblock In {\em CP}, pages 440--448, 2014.

\bibitem{Coq88}
T.~Coquand and G.~P. Huet.
\newblock The calculus of constructions.
\newblock {\em Inf.\ Comput.}, 76(2/3):95--120, 1988.

\bibitem{grit}
L.~Cruz-Filipe and P.~Schneider-Kamp.
\newblock Grit format, formalization, and checkers.
\newblock Available from: \url{http://imada.sdu.dk/~petersk/grit/}.
\newblock Source codes also available from:
  \url{https://github.com/peter-sk/grit}.

\bibitem{lcf:psk:15a}
L.~Cruz-Filipe and P.~Schneider-Kamp.
\newblock Formalizing size-optimal sorting networks: Extracting a certified
  proof checker.
\newblock In {\em {ITP}}, pages 154--169, 2015.

\bibitem{lcf:psk:15b}
L.~Cruz-Filipe and P.~Schneider-Kamp.
\newblock Optimizing a certified proof checker for a large-scale
  computer-generated proof.
\newblock In {\em {CICM}}, pages 55--70, 2015.

\bibitem{dfms-coq09}
A.~Darbari, B.~Fischer, and J.~Marques-Silva.
\newblock Formalizing a {SAT} proof checker in {Coq}.
\newblock In {\em First Coq Workshop, published as technical report tum-i0919
  of the Technical University of Munich}, 2009.

\bibitem{dfms-ictac10}
A.~Darbari, B.~Fischer, and J.~Marques{-}Silva.
\newblock Industrial-strength certified {SAT} solving through verified {SAT}
  proof checking.
\newblock In {\em ICTAC}, pages 260--274, 2010.

\bibitem{goldberg-date03}
E.~I. Goldberg and Y.~Novikov.
\newblock Verification of proofs of unsatisfiability for {CNF} formulas.
\newblock In {\em DATE}, pages 10886--10891, 2003.

\bibitem{drat-trim}
M.~Heule.
\newblock The {DRAT} format and {DRAT}-trim checker.
\newblock CoRR, abs/1610.06229, 2016.
\newblock Source code available from:
  \url{https://github.com/marijnheule/drat-trim}.

\bibitem{heule-appa14}
M.~Heule and A.~Biere.
\newblock Proofs for satisfiability problems.
\newblock In {\em All about Proofs, Proofs for All (APPA)}, July 2014.
\newblock \url{http://www.easychair.org/smart-program/VSL2014/APPA-index.html}.

\bibitem{heule-fmcad13}
M.~Heule, W.~A. {Hunt Jr.}, and N.~Wetzler.
\newblock Trimming while checking clausal proofs.
\newblock In {\em FMCAD}, pages 181--188, 2013.

\bibitem{heule-cade13}
M.~Heule, W.~A. {Hunt Jr.}, and N.~Wetzler.
\newblock Verifying refutations with extended resolution.
\newblock In {\em CADE}, pages 345--359, 2013.

\bibitem{heule-stvr14}
M.~Heule, W.~A. {Hunt Jr.}, and N.~Wetzler.
\newblock Bridging the gap between easy generation and efficient verification
  of unsatisfiability proofs.
\newblock {\em Softw. Test., Verif. Reliab.}, 24(8):593--607, 2014.

\bibitem{heule-cade15}
M.~Heule, W.~A. {Hunt Jr.}, and N.~Wetzler.
\newblock Expressing symmetry breaking in {DRAT} proofs.
\newblock In {\em CADE}, pages 591--606, 2015.

\bibitem{heule-sat16a}
M.~Heule, O.~Kullmann, and V.~W. Marek.
\newblock Solving and verifying the boolean pythagorean triples problem via
  cube-and-conquer.
\newblock In {\em SAT}, pages 228--245, 2016.

\bibitem{heule-fmcad14}
M.~Heule, M.~Seidl, and A.~Biere.
\newblock Efficient extraction of skolem functions from {QRAT} proofs.
\newblock In {\em FMCAD}, pages 107--114, 2014.

\bibitem{biere-sat07}
T.~Jussila, A.~Biere, C.~Sinz, D.~Kr{\"{o}}ning, and C.~M. Wintersteiger.
\newblock A first step towards a unified proof checker for {QBF}.
\newblock In {\em SAT}, pages 201--214, 2007.

\bibitem{biere-sat06}
T.~Jussila, C.~Sinz, and A.~Biere.
\newblock Extended resolution proofs for symbolic {SAT} solving with
  quantification.
\newblock In {\em SAT}, pages 54--60, 2006.

\bibitem{konev-corr14b}
B.~Konev and A.~Lisitsa.
\newblock Computer-aided proof of {E}rd{\H{o}}s discrepancy properties.
\newblock {\em CoRR}, abs/1405.3097, 2014.

\bibitem{konev-sat14}
B.~Konev and A.~Lisitsa.
\newblock A {SAT} attack on the {E}rd{\H{o}}s discrepancy conjecture.
\newblock In {\em SAT}, pages 219--226, 2014.

\bibitem{konev-aij15}
B.~Konev and A.~Lisitsa.
\newblock Computer-aided proof of {E}rd{\H{o}}s discrepancy properties.
\newblock {\em Artif. Intell.}, 224:103--118, 2015.

\bibitem{Letouzey2008}
P.~Letouzey.
\newblock Extraction in {Coq}: An overview.
\newblock In {\em CiE 2008}, volume 5028 of {\em LNCS}, pages 359--369.
  Springer, 2008.

\bibitem{maric-tcs10}
F.~Maric.
\newblock Formal verification of a modern {SAT} solver by shallow embedding
  into {Isabelle/HOL}.
\newblock {\em Theor. Comput. Sci.}, 411(50):4333--4356, 2010.

\bibitem{maric-lmcs11}
F.~Maric and P.~Janicic.
\newblock Formalization of abstract state transition systems for {SAT}.
\newblock {\em Logical Methods in Computer Science}, 7(3), 2011.

\bibitem{mehlhorn-csr11}
R.~M. McConnell, K.~Mehlhorn, S.~N{\"{a}}her, and P.~Schweitzer.
\newblock Certifying algorithms.
\newblock {\em Computer Science Review}, 5(2):119--161, 2011.

\bibitem{shankar-atva08}
N.~Shankar.
\newblock Trust and automation in verification tools.
\newblock In {\em ATVA}, pages 4--17, 2008.

\bibitem{biere-csr06}
C.~Sinz and A.~Biere.
\newblock Extended resolution proofs for conjoining {BDDs}.
\newblock In {\em CSR}, pages 600--611, 2006.

\bibitem{smith-tr08}
D.~R. Smith and S.~J. Westfold.
\newblock Synthesis of satisfiability solvers.
\newblock Technical report, Kestrel Institute, April 2008.

\bibitem{vangelder-isaim08}
A.~Van~Gelder.
\newblock Verifying {RUP} proofs of propositional unsatisfiability.
\newblock In {\em ISAIM}, 2008.

\bibitem{vangelder-sat09}
A.~Van~Gelder.
\newblock Improved conflict-clause minimization leads to improved propositional
  proof traces.
\newblock In {\em SAT}, pages 141--146, 2009.

\bibitem{vangelder-amai12}
A.~Van~Gelder.
\newblock Producing and verifying extremely large propositional refutations -
  have your cake and eat it too.
\newblock {\em Ann. Math. Artif. Intell.}, 65(4):329--372, 2012.

\bibitem{weber-jal09}
T.~Weber and H.~Amjad.
\newblock Efficiently checking propositional refutations in {HOL} theorem
  provers.
\newblock {\em J. Applied Logic}, 7(1):26--40, 2009.

\bibitem{heule-itp13}
N.~Wetzler, M.~Heule, and W.~A. {Hunt Jr.}
\newblock Mechanical verification of {SAT} refutations with extended
  resolution.
\newblock In {\em ITP}, pages 229--244, 2013.

\bibitem{heule-sat14}
N.~Wetzler, M.~Heule, and W.~A. {Hunt Jr.}
\newblock {DRAT}-trim: Efficient checking and trimming using expressive clausal
  proofs.
\newblock In {\em SAT}, pages 422--429, 2014.

\bibitem{wetzler-phd15}
N.~D. Wetzler.
\newblock {\em Efficient, mechanically-verified validation of satisfiability
  solvers}.
\newblock PhD thesis, The University of Texas at Austin, 2015.

\bibitem{malik-date03}
L.~Zhang and S.~Malik.
\newblock Validating {SAT} solvers using an independent resolution-based
  checker: Practical implementations and other applications.
\newblock In {\em DATE}, pages 10880--10885, 2003.

\end{thebibliography}

\end{document}